\documentclass[letter,referee]{aa}
\usepackage[varg]{txfonts}
\usepackage{graphicx}
\usepackage{natbib}
\usepackage{hyperref}     

\def\grb {GRB\,131108A}

\begin{document}

\title{A prompt extra component in the high energy spectrum of  GRB 131108A}

\author{A. Giuliani \inst{1}
   \and S. Mereghetti\inst{1}
   \and M. Marisaldi \inst{2}
   \and F. Longo \inst{3}
   \and E. Del Monte \inst{4}
   \and C. Pittori\inst{5,6}
   \and F. Verrecchia\inst{5,6}
   \and M. Tavani\inst{4}
   \and P. Cattaneo\inst{7}
   \and L. Pacciani\inst{4}
   \and S. Vercellone\inst{8}
   \and A. Rappoldi\inst{7}
        }

\offprints{A. Giuliani, \email{giuliani@iasf-milano.inaf.it}}

\institute{INAF/IASF Milano, via E.\ Bassini 15, I-20133 Milano, Italy   
            \and INAF/IASF Bologna, via Gobetti 101, I-40129 Bologna, Italy  
            \and Dipartimento di Fisica and INFN Trieste, via Valerio 2, I-34127 Trieste, Italy
            \and INAF/IAPS via Fosso del Cavaliere 100, I-00133 Roma, Italy 
	    \and ASI Science Data Center, via del Politecnico snc, I-00133 Roma, Italy
	    \and INAF/OAR, Via Frascati 33, 00040 Monte Porzio Catone (RM), Italy
	    \and Dipartimento di Fisica and INFN Pavia, via Bassi 6, I-27100 Pavia, Italy
            \and INAF/IASF Palermo, via La Malfa 153, I-90146 Palermo, Italy  
} 

\date{Received   / Accepted }

\abstract{The high-fluence  \grb\ at redshift z=2.4, was detected by the Mini-Calorimeter (MCAL, 0.35--100 MeV)   and the Gamma-Ray Imaging Detector (GRID,  30 MeV -- 30 GeV) onboard the AGILE satellite.
The burst emission consisted of a very  bright initial peak, lasting $\sim$0.1 s, followed by a fainter emission detected for  $\sim$25 s  with the MCAL and $\sim$80 s with the GRID. 
The AGILE spectra, when compared with those reported at lower energies, indicate the presence of a prominent high-energy component with peak energy in the $\sim$10--20 MeV region. 
Contrary to other GRBs, this high-energy component is present also during the initial peak,  with  power law photon index of about --1.6 below 10 MeV and --2.35$\pm$0.2  above 30 MeV.   
} 

\maketitle 

\section{Introduction}

\begin{figure}
\centering
\resizebox{8cm}{8.4cm}{\includegraphics{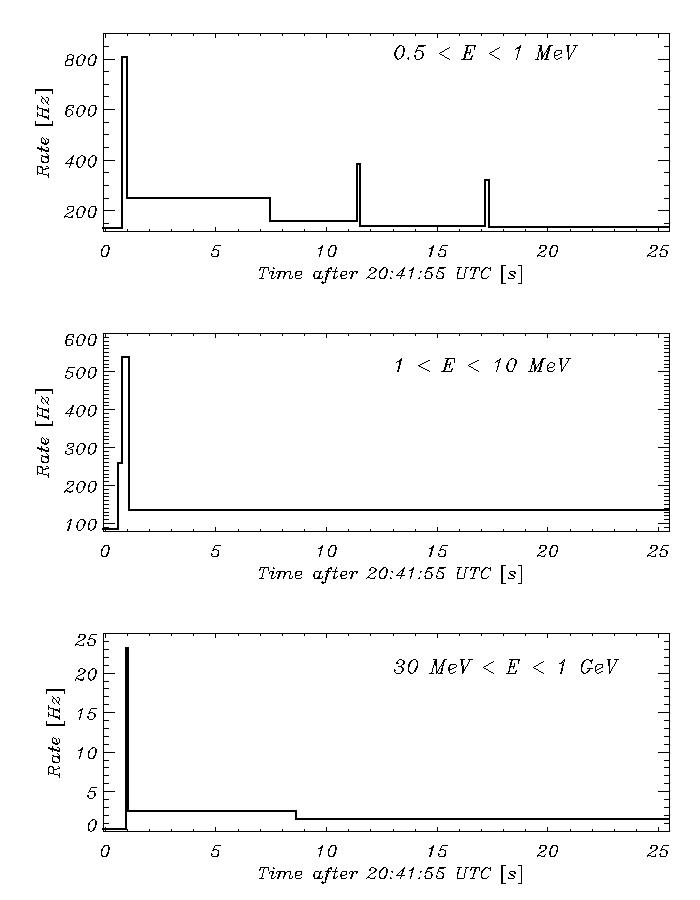}}        
\caption{Light curves of \grb\ obtained with the MCAL,  in the 0.5--1 MeV (top panel) and  1--10 MeV (middle panel) energy range, and with the GRID in the  30 MeV --1 GeV range  (bottom panel). The binning has been optimized using the Bayesan block method (see text). }
\label{fig:lcmcal}
\end{figure}

Despite the large number of observations obtained in the last few years by the Fermi and AGILE satellites, the emission of $\gamma$-ray bursts (GRBs) in the  high-energy  band (E > 50 MeV) is still difficult to interpret.
A typical property of the GRBs emission in this band is its  long duration, which can be one order of magnitude longer than that observed in the  hard X-ray band. 
Moreover, the onset of the high-energy emission is often  delayed with respect to that at lower energies. 
On longer time scales,  the high-energy $\gamma$-ray light-curves decay with time  as a power-law  t$^{-a}$, with $0.9 < a < 1.6$ \citep{ackermann13}.  
Such a power-law decay is similar to that of  the afterglow observed at longer wavelengths and suggests an external shock origin also for the high-energy $\gamma$-rays. 
\\
Also from the spectral point of view the high-energy $\gamma$-ray emission shows some peculiarities. 
AGILE and Fermi observations have shown  that  several bursts exhibit a significant deviation from the Band law \citep{band93}, which was previously believed  to give an adequate description of GRB emission also at high energies.
In fact, while for some GRBs the spectrum of the prompt emission can be fitted by a single  Band function from tens of keV up to tens of GeV, in most cases  the presence of  an additional  hard component at energies above some tens of MeV is required \citep{zhang11}.
\\
The Gamma-Ray Imaging Detector \citep[GRID,][]{tavani09} on board the AGILE satellite is  particularly suited for the observation of GRBs thanks to its large field of view (FOV), which covers one fifth of the sky in the 30 MeV -- 30 GeV energy range, with a very short  deadtime of   $\sim$100 $\mu$s.
The AGILE Mini-Calorimeter (MCAL), besides being exploited as part of the GRID, can be used to autonomously detect and study GRBs in the 0.35--100 MeV range with excellent timing \citep{labanti09,marisaldi08}.
\\
AGILE has contributed to increase the number of high-energy observations revealing, so far, 6 GRBs  with emission above
30 MeV \citep{giuliani08, giuliani10, delmonte11, longo12}. 
Here we present the results obtained for the very energetic  \grb\  with both the GRID and the MCAL instruments.

\section{GRB 131108A}
\label{sec:grb}

\grb\  was discovered by the Fermi/LAT instrument on November 8th, 2013  at  20:41:55.76 UTC   \citep{GCN15464}. 
The burst spectrum at  E $>$ 100 MeV is consistent  with a power law with photon index --2.66$\pm$0.12 and an average flux of  (10$\pm$1.3)$\times$10$^{-9}$ erg cm$^{-2}$ s$^{-1}$
during the firsts 1500 s  after the trigger time \citep{GCN15472} \footnote{see also : \textit{http://www.asdc.asi.it/grblat}}.
\\
At lower energies, \grb\ was detected with the Fermi Gamma-Ray Burst Monitor (GBM)  in the range 10--1000 keV \citep{GCN15477} and with the  Konus/Wind  instrument in the range 20 keV--10 MeV \citep{GCN15480}. The spectra obtained by these two instruments are consistent with a Band function, with parameters E$_{peak} = 373 \pm 14$ keV, $\alpha = -0.91 \pm 0.02$, and $\beta = -2.6 \pm 0.1$ for the Fermi/GBM and E$_{peak} = 340 \pm 11$ keV, $\alpha = -1.11 \pm 0.05$ and $\beta = -2.72 \pm 0.06$ for Konus/Wind.
The fluences measured by these two instruments were (3.65$\pm0.04)\times10^{-5}$ erg cm$^{-2}$ (Fermi/GBM) and (4.15$\pm0.25)\times10^{-5}$ erg cm$^{-2}$ (Konus/Wind).
\\
The Swift satellite started to observe the Fermi error region of \grb\ 65 minutes after the trigger time  \citep{GCN15465, GCN15468, GCN15474} and discovered the   X-ray afterglow, with flux decaying as  $\sim$t$^{-a}$, with $a$=1.9$^{+0.7}_{-0.6}$,   at the coordinates R.A.= 10$^{\rm h}$ 26$^{\rm m}$ 0.43$^{\rm s}$, Dec.=+09$^{\circ}$ 39$'$ 44.9$"$.
\\
A bright optical counterpart, with visual magnitude of 18.5, was found  with the CAHA telescope about 7 hr after the GRB trigger\citep{GCN15469}. Shortly later,  an optical spectrum was obtained with the GTC telescope at the  Roque de los Muchachos  Observatory. Several absorption lines corresponding to Lyman-alpha and highly ionized metals were identified, giving a  redshift  $z$=2.40  \citep{GCN15470, GCN15471}.  The optical afterglow rapidly faded  below magnitude 20 in the following hours \citep{GCN15482, GCN15484}.
At radio wavelengths a source consistent with the location of \grb\ was detected by  the Very Large Array, with flux $\sim$200 $\mu$Jy at 19 GHz about 2.6 days after the burst \citep{GCN15502}.

\section{Data analysis}
\label{sec:agile}

The events collected by the AGILE/GRID detector are reconstructed and processed on board in order to reduce the Earth albedo photons and the particle background in the data transmitted to ground \citep{cocco02,longo02,giuliani06}.
The downloaded  data are then reprocessed by the on-ground analysis software, which assigns to each event an arrival direction, an energy, and a flag related to the probability that the event was produced either by a gamma photon or a charged particle. 
The events are hence divided into  four classes: P (particles), L (limbo), S (single tracks) and G (gammas) in which the 
residual contamination of the background is progressively reduced \citep{bulgarelli10,bulgarelli12}. 
Similarly to what was done for other GRBs observed by AGILE \citep{giuliani08, giuliani10, delmonte11}, for the spectral and light curve analysis of \grb\ the events of the L, S and G classes were used.
This allows us to obtain an effective area larger than that used in the  standard analysis of other $\gamma$-ray sources (based on G events only), especially at low energies. In the case of GRBs, the large source flux  during short time intervals makes the  background negligible compared to the signal.
\\
Since November 2009 AGILE is observing the $\gamma$-ray sky in ``spinning mode''. 
In this operating mode, the satellite rotates around an axis perpendicular to the GRID field of view direction with a period of $\sim$7 minutes.
The sky position of \grb\  at the time of its onset was in the GRID FOV,
at an off-axis angle of $\sim$40$^{\circ}$, and it crossed the FOV during the following 110 s.  
The smallest off-axis angle, $\sim$15$^{\circ}$, was reached  at $\sim$T$_0$+50 s (here and in the following we define  T$_0$ = 20:41:55 UTC ).
During this first transit, the GRID  detected  71 events (L+S+G class) with arrival direction within 20$^{\circ}$ from the burst position. 
This  corresponds to an excess of 5.5 $\sigma$ above the background level \citep{GCN15479}.
In the following rotations of the satellite, the GRB region was observed again with the GRID several times, but no significant emission from the position of \grb\ was detected. 
\\MCAL triggered on \grb\ and delivered, for every photon detected in the time interval [ T$_0$--1.4 s, T$_0$+32.8 s], energy information in the 0.35 - 100 MeV range and a time tag with 2 microsec accuracy.

\begin{figure}
  \resizebox{\hsize}{!}{\includegraphics[height=12cm]{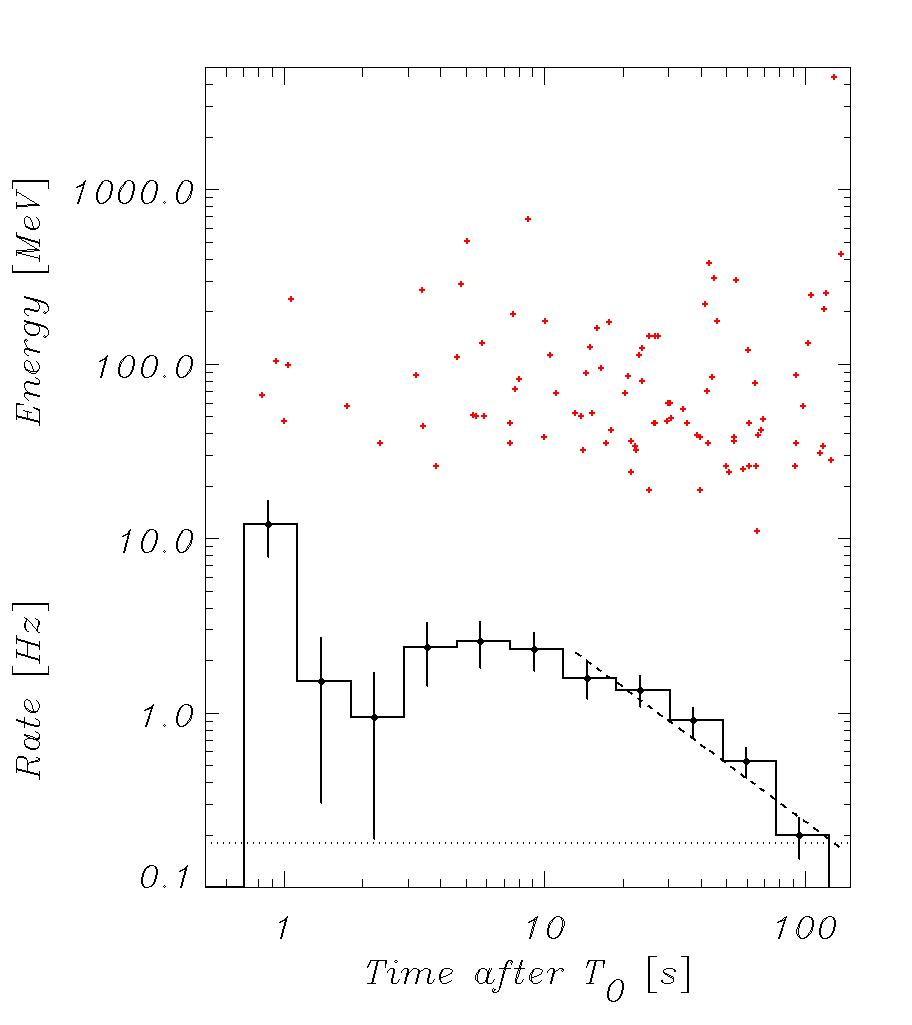}}
  \caption{The lower part of the figure shows the GRID light curve for \grb\ in logarithmic time bins. The dotted line marks the background level and the dashed line indicates a decaying law proportional to $t^{-1.1}$. The upper part of the plot (red points) shows the  energies of the GRID photons associated to the GRB.}
  \label{fig:lc}
\end{figure}


\section{Results}
\label{sec:results}

\begin{figure*}
\centering
  \includegraphics[width=15cm]{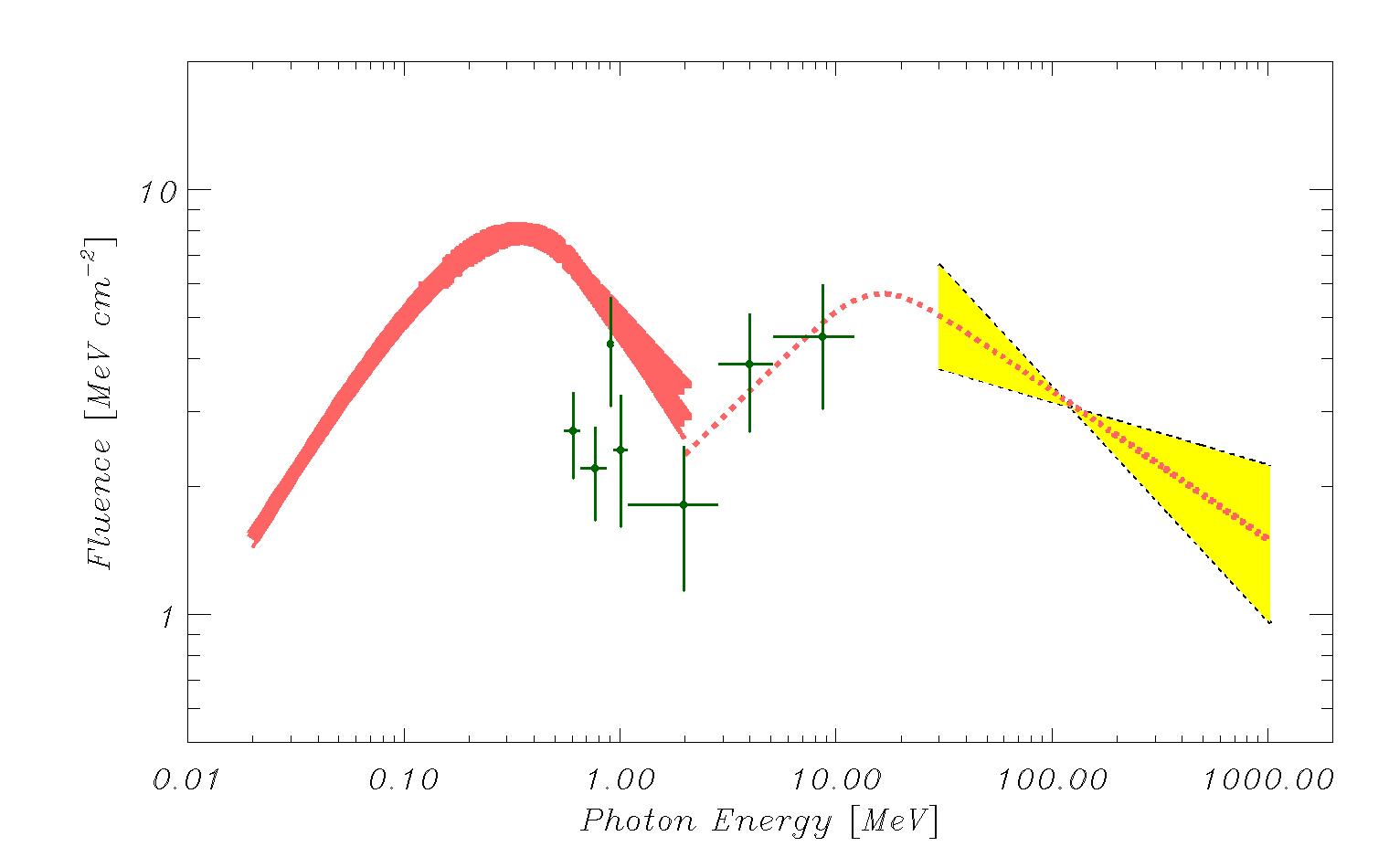}
     \caption{Spectral energy distribution of the \grb\   emission  from T$_0$ to T$_0$ + 25 s.  AGILE/MCAL: crosses, AGILE/GRID: yellow area,  
     Konus/Wind: red belt (from \cite{GCN15480}). The red dotted line represents a smoothly broken power-law with break at $E_b=$15 MeV and indices $\alpha=$--1.5 and $\beta=$--2.35 }
     \label{fig:sed}
\end{figure*}

The light curves of \grb\ obtained with the MCAL and the  GRID  are shown in figure \ref{fig:lcmcal} for different energy bands.
The binning of these light curves was determined using the Bayesan block method \citep{scargle13}, which provides an objective representation of the data with a minimum number of a priori assumptions. \grb\ is characterized by a short ($\sim$0.1 s) initial spike, visible at all energies, followed by a long tail of emission, which in the lower MCAL range remains above the background level for about 25 s.  
\\
We first extracted a spectrum from the MCAL data in the time interval from T$_0$  to T$_0 + 25$ s. A power-law fit  in the energy range 0.5$<$ E $<$ 10 MeV, gives a photon index of --1.90$\pm$0.13 and a fluence of  (1.72$\pm$0.36)$\times$10$^{-5}$ erg cm$^{-2}$ in the range 0.4 - 20 MeV ($\chi^2_r = 1.467$,  corresponding to a null hypothesis probability of 0.11).
\\
We then examined the MCAL spectra corresponding to the initial pulse (from T$_0$+0.79 s to T$_0$ + 1.08 s) and to the tail  (from T$_0$ + 1.08 s to T$_0 + 25$ s).
Power law fits gave a photon index of --1.54$\pm$0.08 ($\chi^2_r =  0.404$) for the pulse and of --1.71$\pm$0.11 ($\chi^2_r = 1.055$) for the tail.
\\
Figure \ref{fig:lc} shows the GRID light curve in logarthimic time bins, obtained from the  events with arrival directions  within 20$^{\circ}$  from the position of \grb .  
The energies of the individual photons  are plotted in the upper part of the same figure. 
%
After the bright initial peak,  the GRID flux remained nearly constant  for $\sim$20 s 
and then decreased,   reaching  the background level (dotted line) after $\sim$ 80 s.
The GRID light curve after T$_0$+20 s can be roughly fitted by a power-law  F(t) $\propto$ t$^{-a}$ with a = 1.1.
The fluence  in the  30 MeV -- 1 GeV range  from T$_0$ to T$_0$+80 s is (2.02$\pm$0.25)$\times10^{-5}$ erg cm$^{-2}$.
\\
The small number of GRID counts does not allow us to carry out a detailed spectral analysis, but we can infer some spectral information from the observed ratio of counts below and above 100 MeV. This ratio is about 4:1, indicating a  spectrum  softer than that of the other GRBs observed by AGILE,  for which the same selection of events gave a 1:1 ratio  \citep[see e.g.][]{giuliani10}.
Assuming a power-law spectrum, and taking into account the instrumental response,
the above ratio for \grb\ corresponds to a photon index  of --2.35$\pm$0.2.
There is no evidence of changes in the spectrum during the GRB.

\section{Discussion}
\label{sec:discussion}

Figure \ref{fig:sed} shows the combined MCAL plus GRID spectrum of the GRB emission from T$_0$ to T$_0$ + 25 s (the GRID flux given above has been rescaled proportionally to the number of events observed in this time interval assuming the same spectral slope measured for the entire observation).
\begin{figure}
\centering
  \resizebox{\hsize}{10cm}{\includegraphics{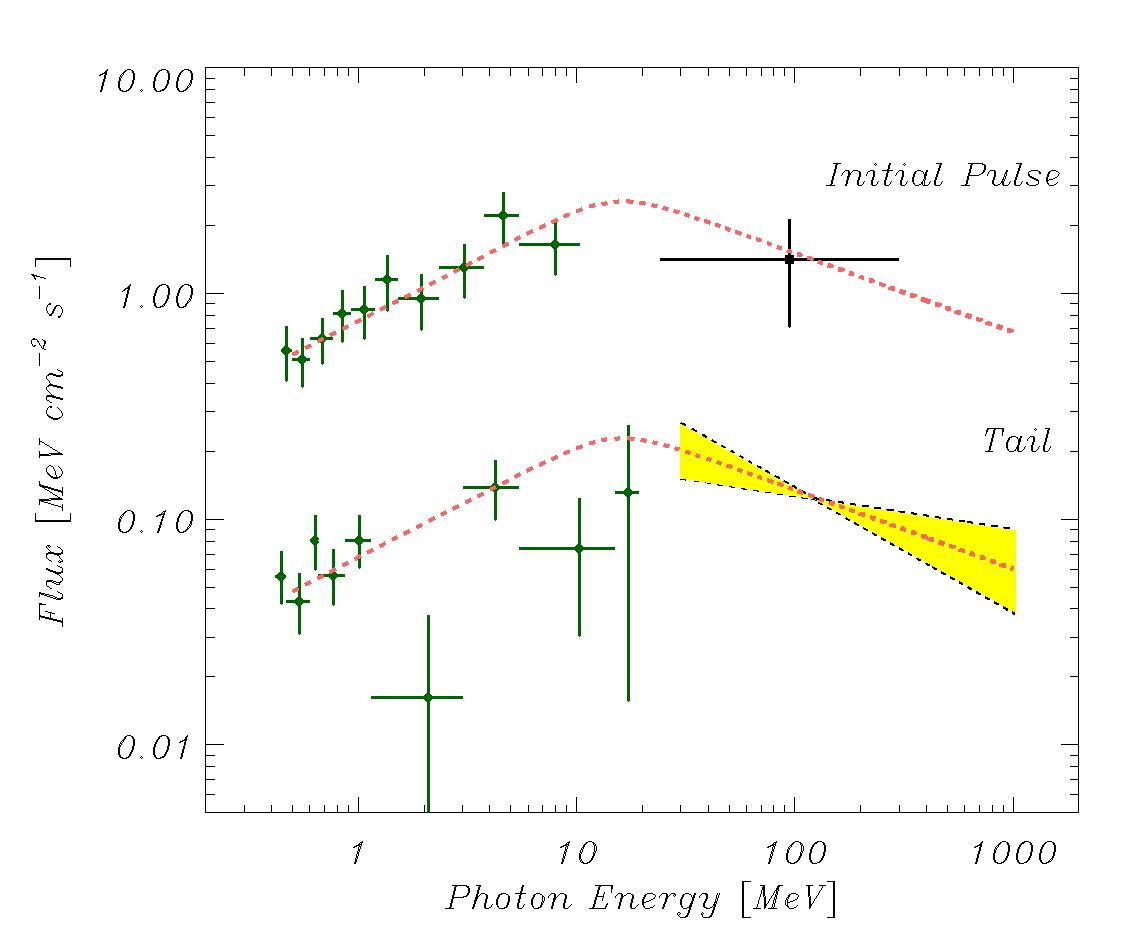}}
     \caption{Average spectra of  \grb\ during the initial pulse (T$_0$ + 0.79 s to T$_0$ + 1.08 s, upper plot) and in the following tail (T$_0$ + 1.08 s to T$_0 + 25$ s, lower plot), as seen by MCAL (green crosses) and GRID instruments (black cross and yellow area).}
     \label{fig:sed2}
\end{figure}
%
 
%
%
For comparison we plot in the same figure the Konus/WIND spectrum  for  
approximately the same time interval as  reported by \citet{GCN15480}.
The AGILE measurements are inconsistent with an extrapolation of the Konus-Wind data, indicating the presence of an additional spectral component 
above a few MeV and with a broad peak between 10 and 20 MeV. 
The MCAL and GRID spectra can be connected by a smoothly broken power-law (dotted line in figure \ref{fig:sed}) with $E_b=$15 MeV and indices $\alpha=$--1.5 and $\beta=$--2.35. 
\\
To better investigate the origin of this high-energy component, we consider separately the spectra of the initial pulse.
This is done  in figure \ref{fig:sed2}, which shows the MCAL spectrum from T$_0$ +  0.79 s to T$_0$ + 1.08 s and  the GRID flux obtained from the 5 photons detected in the same time interval. The error bar of the GRID flux includes, besides the statistical error, the systematics due to the uncertainties in the photon energies and effective area. The MCAL data indicate that the extra-component peaking in the $\sim$10--20 MeV region,  is already present during the initial pulse.
The lower spectra in Fig.\ref{fig:sed2} correspond to the burst tail (T$_0$ + 1.08 to T$_0 + 25$ s). 
\\
An extra component in the high-energy spectrum has been already observed in other GRBs,  such as  080916C, 090510, 090902B, 090926A and 110731A \citep{ackermann13, giuliani10}.  
However in these bursts the extra component showed a delayed onset with respect to the low energy component described by the Band component. 
This behaviour is naturally explained in the early afterglow models \citep{kumar09,ghisellini10,depasquale10,razzaque10}  predicting that the high-energy emission is generated in the external shock. 
\\
However, for \grb , the similarity of the light curves and the lack of a delay in the extra component suggest that the two spectral components in this case are related and arise from the same emission site.
One possibility is that the extra component in the spectrum is produced by Compton upscattering of the Band component \citep[see e.g.][]{bosnjak09,toma11}. 
Alternatively it could be due to synchrotron emission from hadrons \citep{asano09,razzaque10b}.

% ...
%
%
% ...

\section{Conclusions}
\label{sec:conclusion}

In a few respects the properties of \grb\ are in line with those of other GRBs detected above a few tens of MeV:
it had a large fluence in the hard X-ray/soft $\gamma$-ray range and its high-energy emission decayed with time as a power law.   The AGILE data reported give evidence of the presence  of a spectral component peaking  in the $\sim$10--20 MeV region  with  fluence of 3.3$\times$10$^{-5}$ erg  cm$^{-2}$  (E > 2 MeV), in addition to that seen below 10 MeV and  well fit by a Band function in the data of other satellites.
\\
The high redshift of \grb\ (z=2.4) makes it one of the most distant GRBs seen in high-energy $\gamma$-rays.  Furthermore, its large fluence (in the top 10\% of  the bursts in the Fermi/GBM  catalog \citep{goldstein12}) implies a large intrinsic energy.  Assuming standard cosmological parameters ($H_0= 71$, $\Omega_m = 0.27$, $\Omega = 1$) its isotropic energies  are  ($4.24\pm0.05)\times10^{53}$ erg, for the  Band component,  and $(2.35\pm0.29)\times10^{53}$ erg for the additional spectral component found by AGILE.
The high value of the fluence and the isotropic luminosity put this event in the small class of hyperfluent GRBs. 
\\It is also interesting to note that the prominent spectral component detected by AGILE above the observed energy of 10 MeV is peaking in the burst rest frame at $E_{rf}^{peak} = (1 + z) \, E_b \simeq 50$ MeV. Modelling this extra component is of the greatest importance, and it will be the subject of forthcoming investigations.

%




%

\bibliographystyle{aa}
\bibliography{agile,gcn,grb}

\end{document}